\documentclass[conference]{IEEEtran}
\IEEEoverridecommandlockouts
\usepackage{amsmath}
\usepackage{array}
\usepackage{booktabs}
\usepackage{tabularx}
\usepackage{cite}
\usepackage{url}
\usepackage{graphicx}

\newcolumntype{Y}{>{\raggedright\arraybackslash}X}
\newcommand{\MusMet}{MM}

\usepackage{tikz}
\usetikzlibrary{positioning,arrows.meta,fit,calc}
\begin{document}

\title{When Music Needs Meaning: Semantic Communication as a Transformative Enabler for the Musical Metaverse\thanks{This work is supported by MUSMET project funded by the EIC Pathfinder Open scheme of the European Commission (grant agreement n. 101184379), and has received funding from the
Swiss State Secretariat for Education, Research and Innovation (SERI). Corresponding author: ali.alhousseini@supsi.ch}}

\author{  
\IEEEauthorblockN{
    Ali Al Housseini\IEEEauthorrefmark{3},
    Jaime Llorca\IEEEauthorrefmark{2}\IEEEauthorrefmark{4}\IEEEauthorrefmark{5},
    Cristina Rottondi\IEEEauthorrefmark{1},
    Luca Turchet\IEEEauthorrefmark{2},
    Tiziano Leidi\IEEEauthorrefmark{3},
    Omran Ayoub\IEEEauthorrefmark{3}\\
    }

  \IEEEauthorblockA{\IEEEauthorrefmark{1}
    Politecnico di Torino, Turin, Italy, \IEEEauthorrefmark{2}
    University of Trento, Trento, Italy}
    \IEEEauthorblockA{\IEEEauthorrefmark{3}
    University of Applied Sciences and Arts of Southern Switzerland, Lugano, Switzerland}
    \IEEEauthorblockA{\IEEEauthorrefmark{4}
    Centre Tecnologic de Telecomunicacions de Catalunya (CTTC/CERCA), Castelldefels, Spain.}
    \IEEEauthorblockA{\IEEEauthorrefmark{5}
    New York University, New York, USA.}

  \small ali.alhousseini@supsi.ch
}

\maketitle

\begin{abstract}
The Musical Metaverse (\MusMet{}) enables geographically distributed musicians, audiences, sound engineers, and AI agents to interact through synchronized audio, visual, kinematic, and haptic streams, imposing latency, synchronization, bandwidth, and scalability requirements that are difficult to satisfy with conventional bit-level communication. This paper argues that semantic communication, understood here as the transmission of musically meaningful information rather than raw audio-visual-haptic samples, provides a principled communication paradigm for the \MusMet{}. We define the \MusMet{} from a communication perspective and derive modality-specific requirements; propose a three-pillar framework for musical semantic communication comprising semantics extraction, semantic-aware orchestration, and cross-modal semantic synchronization; survey representative work across semantic communication, music AI and edge orchestration, identifying the gaps that prevent their integration; and propose a semantic-first reference architecture based on device-side semantic encoders, edge semantic routers, and cross-modal semantic synchronization. The resulting framework positions the \MusMet{} as a concrete and demanding use case for 6G semantic communication.
\end{abstract}
 
 
\section{Introduction}
 
Networked music performance (NMP) has reached a point at which geographically separated musicians can, under favorable network conditions, play together with end-to-end audio latency in the range of 30--40 ms \cite{rottondi2016}. Systems such as JackTrip, Soundjack, and Jamulus, combined with advances in 5G slicing, edge computing, and optimized codecs, have made point-to-point remote performance increasingly practical \cite{gabrielli2025,rottondi2025,boem2025}. However, the Musical Metaverse (\MusMet{}) demands more than low-latency audio between two endpoints \cite{rottondi2016,turchet2024}. It requires a distributed, real-time, multi-user, and multi-modal environment in which musicians, audiences, sound engineers, and AI agents interact through audio, motion, haptic feedback, visual scenes, and semantic metadata \cite{boem2025,aloudat2025}.
 
The challenge extends beyond transporting large volumes of heterogeneous data. In an NMP, not all information contributes equally to the shared musical experience. In other words, different data streams carry different semantic roles and therefore require different communication priorities. For instance, the onset of a drum hit or a conductor's gesture may be essential for rhythmic synchronization. In contrast, the fine detail of a reverberation tail or the resolution of a background visual texture can often tolerate degradation or delay. Similarly, closed-loop haptic interaction may demand sub-10~ms responsiveness~\cite{turchet2024}, whereas haptic monitoring cues and many visual elements tolerate looser latency constraints. Conventional communication design, following Shannon's paradigm, focuses on reliable and efficient delivery of bits and deliberately abstracts away the meaning of the message. Quality-of-service mechanisms can prioritize flows~\cite{rfc2475}, but they act on traffic classes, not on the musical function of individual events within a stream.
Semantic communication addresses this gap, offering a fundamentally different perspective: it seeks to identify the information most relevant to the receiving task, here preserving musical intent, interaction, and perception, and to allocate communication resources accordingly \cite{nguyen2025,chaccour2025}.

This problem becomes increasingly complex as the performance scales from a small ensemble of musicians, which may combine direct audio monitoring with a limited number of remote streams, to immersive concerts that combine audio, motion capture, haptic feedback, avatar state, visual scenes, and audience interaction. 
In such immersive concerts, every additional participant and modality adds streams that must be delivered, aligned, and adapted to heterogeneous devices. Multicast and relay replication reduce what each source must publish, but they do not determine which information each receiver needs for its musical role. The central question is therefore not only how to lower latency, but how to preserve the information that enables musical coordination when bandwidth, transmission opportunities, and edge processing capacity are limited.

Music is a particularly informative use case for semantic communication because of three properties that distinguish it from generic immersive media:
\begin{itemize}
    \item \textbf{Shared temporal reference} -- musicians and audiences coordinate around beats, bars, phrases, and expressive timing, rather than around the absolute time references maintained by network synchronization protocols \cite{rottondi2016};
    \item \textbf{Perceptual priority hierarchy} -- rhythm and onset timing are often more critical than timbral detail, and harmonic structure can be more important than spatial texture \cite{rottondi2025,cairns2025};
    \item \textbf{Creative adaptation} -- musicians anticipate, compensate, and reshape their behavior in response to network impairments in ways that purely mechanical control systems cannot \cite{rottondi2025,boem2025}.
\end{itemize}
Together, these properties make music a demanding test case: a network that can act on musical meaning should protect what coordination depends on and degrade what it does not. Whether it can do so better than well-engineered conventional baselines is the hypothesis this paper frames.

We argue that semantic communication, understood here as transmitting and acting on musically meaningful information rather than only raw signal samples, is a natural enabling paradigm for the MM. 
We do not assume that a complete musical semantic communication system exists. Rather, its components are emerging in separate communities. 
Neural audio codecs and representation-learning models show that musical structure can be encoded compactly \cite{defossez2022,kumar2023,li2024}; edge-AI orchestration provides a substrate for semantic routing \cite{molner2025,zhang2025semantic}; and cross-modal AI provides early mechanisms for aligning information across sensory modalities \cite{wang2024,mao2025}. 
Existing work on semantic communication for the metaverse~\cite{lyu2025,aloudat2025,bian2025} targets generic immersive media and does not model the musical time base or the task-dependent importance of musical events. The missing contribution is an integrated, music-specific communication framework. The components' maturity is, however, uneven: event-level synchronization across audio, haptic, kinematic, and visual modalities remains unvalidated. We therefore present a framework and reference architecture to be tested incrementally through simulation, prototyping, and field trials, and report no empirical validation here.
 
This paper makes four contributions:
\begin{enumerate}


    \item a communication-oriented definition of the \MusMet{}, with the modality-specific latency, alignment, and rate quantities that a deployment must declare, and an analysis of scale (Section~II);
    \item a selective review of representative work across NMP, semantic communication, music AI, neural audio coding, and edge orchestration, whose integration gaps we organize into a three-pillar framework: semantics extraction, semantic-aware orchestration, and cross-modal semantic synchronization (Section~III);
    \item a taxonomy of musical semantics by modality and abstraction level, with candidate task-level diagnostics for assessing whether declared musical features and output relationships survive transmission (Section~IV);
    \item a semantic-first reference architecture composed of device-side semantic encoders, edge musical priority routing, and a semantic synchronization service, illustrated through a worked musical example and an explicit event and timing contract (Section~V).
\end{enumerate}
Section VI concludes the paper with a staged research roadmap that turns the framework into falsifiable research questions.

\begin{table*}[t]
\caption{Representative \MusMet{} Use Cases. Latency values are indicative targets; see Table~\ref{tab:modality_requirements}.}
\label{tab:use_cases}
\centering
\begin{tabularx}{\textwidth}{@{}p{0.13\textwidth}p{0.17\textwidth}p{0.17\textwidth}p{0.17\textwidth}Y@{}}
\toprule
\textbf{Use case} & \textbf{Participants} & \textbf{Modalities} & \textbf{Latency focus} & \textbf{Semantic opportunity} \\
\midrule
UC1: Remote jamming & 3--5 musicians & Audio, kinematic & 30--40 ms audio; low, stable gesture delay & Protect rhythm-critical events under congestion; degrade secondary detail first. \\
UC2: Virtual concert & 10+ musicians; 1K--10K audience members & Audio, haptic, visual, kinematic & Per-modality targets of Table~\ref{tab:modality_requirements}, at audience scale & Transmit musical meaning once; receivers render to their capabilities (semantic multicast). \\
UC3: AI co-creation & One or more humans and AI agents & Audio, kinematic, semantic metadata & Role-dependent: in-loop vs.\ structural time scales & Exchange intent, structure, and expressive cues rather than waveforms. \\
\bottomrule
\end{tabularx}
\end{table*}


\section{The Musical Metaverse: Definition and Requirements}

This section defines the \MusMet{}, presents three use cases, and identifies task-dependent timing, representation, and scaling requirements.
 
\subsection{Definition and Stakeholders}
 
Building on~\cite{turchet2024,boem2025}, we define the \MusMet{} as a distributed, real-time, multi-user, multi-modal immersive environment in which geographically separated participants interact through audio, kinematic, haptic, and visual streams coordinated around a shared musical context.

Real-time operation, multi-modality, scale, and geographic distribution are shared with other immersive applications. The musical specialization has two aspects. First, coordination is expressed in musical time (beats, phrases, and events), which must be mapped to physical time for output scheduling. Second, the importance of a stream element depends on the musical function it serves for a given receiver and context. The semantic view adopted in this paper exploits both aspects: musical time and function can inform routing, compression, and rendering decisions, subject to task-specific evaluation.
 
Four participant groups directly interact within the MM: 
\begin{itemize}
    \item \textbf{Musicians} -- generate and receive performance streams;
    \item \textbf{Audiences} -- consume and, in interactive settings, influence the experience;
    \item \textbf{Sound engineers} -- operate mixing, monitoring, and production workflows within the performance loop;
    \item \textbf{AI agents} -- in scenarios that include them, act as co-creators, adaptive mixers, or accompanists.
\end{itemize}

These interactions are enabled by platform providers, which operate the underlying network, edge, and cloud infrastructure.
 
\subsection{Motivating Use Cases}
 
Three use cases illustrate the operating range of the \MusMet{}. Table~\ref{tab:use_cases} summarizes their participants, modalities, timing concerns, and opportunities for task-aware information selection.
 
\textit{UC1: Remote jamming.} A small ensemble combines local monitoring with remote performance streams. Tolerable source-to-ear delay depends on the musical task, while delay variation can disrupt coordination \cite{rottondi2016,rottondi2025}. During congestion, an explicit monitoring request could guide protection of particular onsets or other required features; any advantage over conventional priority-aware delivery needs evaluation.
 
\textit{UC2: Virtual concert.} A distributed ensemble performs live for a larger audience. Musicians retain UC1-like interaction needs; audience playback may accept more buffering. Conventional multicast or relay fan-out can distribute a limited source set. The semantic hypothesis is that shared task-relevant representations could additionally support heterogeneous local renderers, provided their fidelity and computation costs are acceptable.
 
\textit{UC3: AI co-creation.} AI agents acting as  an accompanist, structural adviser, and mixing assistant have different information and response-time needs. Some decisions may use note or harmonic descriptors; timbre-sensitive mixing may require acoustic detail or audio. Descriptor sufficiency must therefore be established per task rather than assumed for every AI agent.
 
Together, the use cases show that no single latency target characterizes the \MusMet{}: the binding constraint shifts from inter-musician timing (UC1) to scalable multi-modal distribution (UC2) to role-dependent responsiveness (UC3). 
Because the same stream element can matter differently to different receivers, this role dependence motivates semantic metadata that describes stream content in musical terms, combined with receiver requests that declare the function each receiver needs it for.
 
\subsection{Per-Modality Requirement Specification}

Latency, relative alignment, and representation rate are distinct requirements. A source-to-output budget includes capture, extraction, transport, buffering, and rendering; a network-delay measurement alone does not establish it. Synchronization error is measured between specified output landmarks after subtracting intended offsets, not necessarily between equal timestamps. Table~\ref{tab:modality_requirements} identifies the quantities a deployment must specify, informed by NMP and musical-interaction studies \cite{rottondi2016,turchet2024,cairns2025}; it does not assign universal perceptual thresholds to entire modalities.
For orientation, the literature reports indicative figures such as 30--40~ms for interactive audio~\cite{rottondi2016} and sub-10~ms responsiveness for closed-loop haptic interaction~\cite{turchet2024}; these depend on task, tempo, and instrumentation and are not used as thresholds in this paper.

Fixed delay, jitter, and intermodal skew can have different consequences. Audience buffering cannot demonstrate feasibility for performer interaction or a haptic control loop. Bitrate depends on sample format, codec, sensor dimensionality, metadata, and local rendering; report both payload and control/transport overhead. Neither semantic metadata nor compression can remove propagation or causal observation delay.

\begin{table*}[t]
\caption{Task-Specific Requirements to Declare Before Evaluation. These are design questions, not measured bounds or validated universal thresholds.}
\label{tab:modality_requirements}
\centering
\begin{tabularx}{\textwidth}{@{}p{0.13\textwidth}p{0.27\textwidth}p{0.26\textwidth}Y@{}}
\toprule
\textbf{Modality} & \textbf{Latency scope} & \textbf{Alignment landmark} & \textbf{Rate/fidelity variables} \\
\midrule
Audio & Source-to-ear monitoring or buffered audience playback. & Rendered onset relative to intended musical timing. & Channels, sample format, codec, and requested pitch/timbre detail. \\
Kinematic & Sensor capture to the receiving task or avatar action. & Defined gesture landmark; a gesture may precede sound. & Tracked joints/state variables, update rate, and quantization. \\
Haptic & Distinguish monitoring cues from closed-loop force interaction. & Measured cue onset/contact peak, including actuator delay. & Actuator channels, update rate, and cue versus force fidelity. \\
Visual & Distinguish interactive avatars from buffered concert video. & Displayed contact frame or other declared visual landmark. & Frame rate, resolution, scene complexity, and representation. \\
\bottomrule
\end{tabularx}
\end{table*}

\subsection{The Scaling Problem}

Let $N_s$ sources publish $M$ modality streams to $N_r$ recipients. Separate source-to-recipient unicast requires up to $N_sN_rM$ logical streams; a full mesh is the special case $N(N-1)M$, where $N_s=N_r=N$. With multicast or relay replication, a source can publish once per group/modality, although receiver delivery, fan-out, and adaptation costs remain. This topology benefit also applies to ordinary encoded media \cite{rfc1112}; it is not a semantic-communication gain.

The additional hypothesis concerns what is published and how recipients use it. A shared event representation could reduce required detail for some tasks or enable local cue generation, but may introduce extraction, metadata, and rendering costs. Comparisons must hold source/receiver counts, topology, available information, and total resource budgets constant. A virtual concert with many audience sinks is not a thousand-source full-mesh performance.

Existing differentiated service can enact priorities \cite{rfc2475}. The proposed interface supplies musical fields and receiver requests that inform those priorities. When a conventional scheduler receives the same metadata and executes the same rule, no separate scheduling advantage follows from calling it semantic.

\section{Related Work}
This section illustrates how representative research and protocol specifications supply the proposed pillars. This selective synthesis focuses on the interfaces needed to combine those capabilities for musical interaction.

\subsection{Networked Music Performance}
NMP demonstrates low-latency collaboration under suitable conditions using optimized audio paths, buffering, and network design \cite{rottondi2016,rottondi2025,boem2025,cairns2025}. Peer-to-peer and server-mediated arrangements address different group sizes and monitoring needs. Heterogeneous musical tasks additionally require a shared contract for event identity, content selection, and output relationships, including cross-modal timing and receiver-specific feature preservation. \textbf{Integration question:} how can these task requirements be exposed consistently across otherwise capable transport, mixing, and rendering components?

\subsection{Semantic Communication for the Metaverse}
Relevant literature includes a chapter on metaverse-oriented semantic communication \cite{lyu2025}, surveys of metaverse integration \cite{aloudat2025,bian2025}, and knowledge-driven semantic network design \cite{chaccour2025}. These works address immersive media in general and do not model the musical time base or the task-dependent importance of musical events. The musical setting motivates a more specific interface: polyphonic roles and metrical context, expressive deviations, and the distinction between fixed delay, jitter, and output skew. \textbf{Integration question:} which musical dimensions should a requested task preserve, at what resolution, and under what deadline? The interface specializes existing task-oriented ideas to musical events and receiver functions.

\subsection{Neural Audio Coding and Music Representation Learning}
Neural audio codecs are important precursors to a transmitted musical representation. EnCodec provides streaming neural audio compression \cite{defossez2022}; DAC studies high-fidelity reconstruction \cite{kumar2023}; SoundSpring explicitly addresses loss-resilient learned transmission \cite{yang2025}; MuCodec studies very-low-bitrate music coding \cite{chen2025mucodec}; and OmniCodec investigates semantic-acoustic disentanglement \cite{zhang2026omnicodec}. Such factorization motivates questions about selectable information, but does not by itself establish independent pitch, rhythm, and timbre controls. MERT \cite{li2024} and Stem-JEPA \cite{riou2025} learn music representations; MuLan \cite{huang2022} associates music and language; audio-language models \cite{tang2024} support descriptions and task conditioning. Together, they provide representation and coding components for the proposed interface.

Existing streaming, bandwidth-selection, and resilience mechanisms provide a starting point. Integration requires assessing selectable musical dimensions, the interactive source-to-output budget, and event relationships across modalities together. High real-time throughput does not by itself bound algorithmic look-ahead or startup context. Learned latents may also be difficult to interpret for task selection \cite{li2026xai}.

\textbf{Integration question:} how should an encoder expose musical fields, confidence, causal availability, and rate/fidelity controls that a task-aware receiver can use? Answering this question requires an interface specification and task-specific benchmarks.

\subsection{Edge Orchestration and Resource Allocation}
Edge-AI orchestration provides infrastructure for semantic-aware resource control. AIORA proposes AI-native multi-stakeholder orchestration for 6G \cite{molner2025}. Surveys connect semantic representations with edge computation and collaborative inference \cite{zhang2025semantic,nasif2025}, while foundation-model work formulates joint partitioning and placement under latency and resource constraints \cite{zhang2025partition}. AI-native 6G surveys provide broader architectural context \cite{benzaid2025}. Devices or edges can serve latency-critical tasks; more distant computation is suitable only when the task's deadline permits it. Long observation history alone does not imply tolerance for delivery delay. Closer to the \MusMet{}, SPARQ models queueing-delay-aware resource orchestration for distributed AI applications \cite{spadaccino2026}, and MuMeNet models musical-metaverse service provisioning \cite{alhousseini2025mumenet}.

\textbf{Integration question:} which musical fields, user requests, and uncertainty estimates should feed these resource policies, and how should the resulting decisions be evaluated? Existing queues can enact a supplied policy; the proposed contribution concerns its musical inputs and end-to-end interpretation.

\subsection{Cross-Modal Alignment and Synchronization}
Cross-modal transmission is studied in \cite{wang2024}; broader surveys \cite{wei2025,nguyen2025} and immersive-generation studies \cite{mao2025,chen2025haptics} cover representation association and generated sensory content. Physical output timing requires an additional contract: a preparatory gesture may precede its associated sound, while a haptic contact cue may target the sound onset. Clock discipline supplies the time coordinates for these intended relationships, with achievable accuracy depending on the deployment.

OSC already carries application-defined messages and schedules bundles by timetag, while requiring an external time reference \cite{osc10}. Ableton Link maintains tempo/beat/phase relationships and maps beat positions to time; aligned beat phase need not imply identical absolute bar numbering \cite{abletonlink}. Thus musical timelines, extensible event labels, and scheduled playback have precedents.

\textbf{Integration question:} how should an application profile associate shared musical identity, intended offsets, confidence, and modality-specific failure policies over those primitives? Section V proposes one bounded contract. Its additional value over ordinary scheduled playback remains unvalidated.

\subsection{Synthesis: From Gaps to Three Pillars}
\label{sec:synthesis}

We organize these integration questions into three coupled functions:
\begin{itemize}
    \item \textbf{Pillar 1: Musical semantics extraction} -- expose task-relevant features and context, including availability and uncertainty, through a source-side \emph{semantic encoder};
    \item \textbf{Pillar 2: Semantic-aware orchestration} -- combine musical requests with network and compute state to choose content and delivery policy;
    \item \textbf{Pillar 3: Cross-modal semantic synchronization} -- associate event identities with intended output relationships above clock discipline.
\end{itemize}
Table~\ref{tab:coverage} summarizes the selected literature; Table~\ref{tab:requirements_gap} links requirements to research questions. A common vocabulary lets the pillars share task requirements while keeping replication topology, representation cost, and timing separately evaluable. P1 builds on codecs and feature extraction; P2 builds on established resource control; P3 combines existing timing primitives with a proposed event/state contract. Their joint causal and perceptual performance remains open.

Section IV defines a communication-oriented taxonomy and candidate diagnostics. Section V connects them through an architecture, a worked example, and an explicit synchronization contract.

\begin{table}[t]
\caption{Coverage of the Three Pillars in the Surveyed Literature}
\label{tab:coverage}
\centering
\begin{tabularx}{\columnwidth}{@{}p{0.20\columnwidth}p{0.24\columnwidth}Y@{}}
\toprule
\textbf{Pillar} & \textbf{Representative works} & \textbf{Coverage and main gap} \\
\midrule
P1: Semantics extraction & \cite{defossez2022,kumar2023,yang2025,chen2025mucodec,zhang2026omnicodec,li2024,riou2025,huang2022,tang2024} & Codecs and features exist; assess named musical controls, causal delay, and receiver fidelity jointly. \\
P2: Semantic orchestration & \cite{molner2025,zhang2025semantic,zhang2025partition,nasif2025,benzaid2025,spadaccino2026,alhousseini2025mumenet} & Resource mechanisms exist; specify musical inputs and task-dependent policy evaluation. \\
P3: Semantic synchronization & \cite{wang2024,mao2025,wei2025,nguyen2025,chen2025haptics,osc10,abletonlink} & Association and timing primitives exist; validate the combined event/state/rendering contract. \\
\bottomrule
\end{tabularx}
\end{table}

\begin{table*}[t]
\caption{Requirements Gap and Corresponding Semantic Communication Pillars}
\label{tab:requirements_gap}
\centering
\begin{tabularx}{\textwidth}{@{}p{0.18\textwidth}p{0.25\textwidth}p{0.14\textwidth}Y@{}}
\toprule
\textbf{Requirement (Sec.~II)} & \textbf{Integration question} & \textbf{Pillar} & \textbf{Proposed response to evaluate} \\
\midrule
Low-latency musical audio & Can a representation support the requested task within its full causal budget? & P1 & Measure extraction, communication, and rendering costs together. \\
Cross-modal synchronization & Which outputs belong to the same event, with what intended offsets? & P3 & Supply event identity and timing intent; measure actual output residuals. \\
Large multi-user scale & What is shared across recipients, beyond conventional replication? & P2 & Compare representation and adaptation costs under the same topology. \\
Bandwidth-constrained operation & Which information may be reduced for this receiver's task? & P1 + P2 & Expose selectable detail and requests to an explicit priority policy. \\
Heterogeneous devices & Which local outputs preserve the requested function? & P1 + P2 & Declare renderer capability and evaluate preserved features, not plausibility alone. \\
\bottomrule
\end{tabularx}
\end{table*}

\section{Musical Semantics for Communication}
 
Section III motivates a shared vocabulary for information that communication and rendering components can act upon. This section defines musical semantics for communication, organizes it by modality and abstraction, and proposes task-level diagnostics for selected fidelity, timing, and affective properties.
 
\subsection{Definition and Criteria}
In this paper, \emph{musical semantics for communication} denotes structured musical descriptors and task-relevant relationships, ranging from observable events to inferred context and intent, organized for exchange under real-time constraints. Their musical relevance follows from the receiving task; descriptor agreement alone does not establish preserved intent or perception. We call \emph{semantic extraction} the process that computes such representations from raw sampled streams; the model performing it is the semantic encoder introduced in Section III.
 
This definition is deliberately narrower than the broad musicological notion of musical meaning: it covers only semantics that satisfy three criteria, each imposing a concrete design constraint.
\begin{itemize}
    \item \textbf{Extractability} -- required fields must become available within the task's causal budget. Report analysis delay, accumulated context, update interval, startup/lock-in time, and look-ahead separately; a fast update rate does not imply a short observation window.
    \item \textbf{Communicability} -- a representation should justify its payload, control overhead, and compute cost through resource savings or improved task preservation relative to an appropriate encoded-media baseline. This is a design criterion to test, not an assumed compression gain.
    \item \textbf{Actionability} -- a receiver or orchestrator must have a specified use for the information, such as selecting note detail, generating a cue, or associating modality outputs. Declare any shared context or residual audio required for that decision.
\end{itemize}
 
\subsection{Modalities}
We organize the semantic content of the \MusMet{} into three primary modality groups and a cross-cutting affective--social dimension. This organization reflects our design perspective rather than an established standard; it follows the stream types identified in Section II.
\begin{itemize}
    \item \textbf{Audio-semantic} -- pitch, rhythm, timbre, articulation, stems, and spatial audio objects. Local measurements and context-dependent estimates must be distinguished: beat position can require historical observations even when updated frequently.
    \item \textbf{Kinematic-haptic} -- gesture, posture, force, vibration, and instrument state. Head tracking provides one concrete example of orientation data in immersive NMP \cite{turchet2024}. Gesture, force, and vibration descriptors in the proposed taxonomy require suitable instrumentation and task-specific validation.
    \item \textbf{Visual-spatial} -- scene geometry, avatar pose, facial actions, gaze, and lighting. Capture, estimation, frame scheduling, and display contribute separately to the task's latency.
    \item \textbf{Affective--social} (cross-cutting) -- emotional trajectories, engagement, turn-taking, and interaction cues. Interpretation is context- and participant-dependent, requiring validation beyond descriptor agreement.
\end{itemize}
 
\subsection{Abstraction Levels}
Musical information can also be organized by abstraction level, from the signal itself to expressive intent. We denote the raw sampled streams as level $L_1$: it is the input to semantic extraction rather than a semantic representation, and it is therefore excluded from the taxonomy of communicable semantics, which comprises three levels.
\begin{itemize}
    \item \textbf{$L_2$: feature semantics} -- pitch, onset strength, beat position, instrument activity, and timbral descriptors. Direct observations may support fast loops; estimates such as beat position can retain longer context. Availability must be measured for the chosen method.
    \item \textbf{$L_3$: structural semantics} -- chord progressions, phrase boundaries, rhythmic hierarchy, key changes, and section labels. These require temporal context; recognizing a past transition does not imply predicting an unannounced future one.
    \item \textbf{$L_4$: intent and affective semantics} -- inferred expressive purpose, tension, surprise, and participant affect. A measured onset deviation is an $L_2$ feature; interpreting it as rubato or expressive intent belongs here. Such interpretations require contextual and perceptual validation.
\end{itemize}
The appropriate level therefore depends on the receiver's role and the time scale of its task, a dependence the architecture of Section V exploits by placing extraction at different tiers.
 
Table~\ref{tab:design_space} maps representative descriptors to the proposed taxonomy. Descriptor availability and inference difficulty vary within each cell; the table does not rank deployment readiness. Integration should begin with task-relevant observable features and add context-dependent layers when their cost and reliability are established.

\begin{figure*}[!t]
\centering
\includegraphics[width=\textwidth]{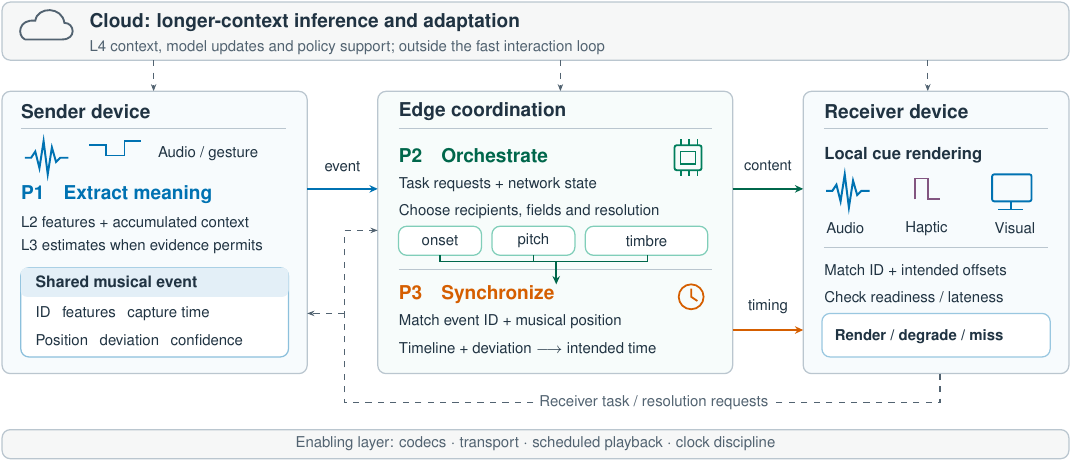}
\caption{Proposed semantic-first architecture. Solid arrows carry event fields, selected content, or timing information; dashed arrows carry model/policy support and receiver requests. A shared event record connects extraction (P1), orchestration (P2), and synchronization (P3). Local features and structural estimates carry confidence; accumulated context is not restricted to a single tier. Conventional codecs, transport, scheduled playback, and clock discipline remain enabling components.}
\label{fig:architecture}
\end{figure*}
 
\begin{table*}[t]
\caption{Design Space of Musical Semantics: Representative Descriptors by Modality Group and Abstraction Level. Implementation readiness is task- and method-dependent.}
\label{tab:design_space}
\centering
\begin{tabularx}{\textwidth}{@{}p{0.16\textwidth}YYY@{}}
\toprule
\textbf{Modality} & \textbf{$L_2$: Feature semantics} & \textbf{$L_3$: Structural semantics} & \textbf{$L_4$: Intent and affect} \\
\midrule
Audio-semantic & Pitch contour, onset strength, beat times, instrument activity. & Chord progression, phrase boundaries, rhythmic hierarchy, key change. & Inferred tension, expressive timing intent, emotional valence, narrative arc. \\
Kinematic-haptic & Gesture velocity, finger pressure, bow position, impact force. & Gesture phrasing, repeated motion patterns, haptic event sequences. & Inferred expressive gesture purpose, anticipated action, perceived effort. \\
Visual-spatial & Avatar joints, facial action units, gaze direction. & Scene transitions, musician proximity, group formation, attention patterns. & Inferred collective emotion, audience engagement, musician--audience rapport. \\
\bottomrule
\end{tabularx}
\end{table*}
 
\subsection{Candidate Task-Level Diagnostics}
Objective quality models such as POLQA (ITU-T P.863) for speech \cite{itup863} and PEAQ (ITU-R BS.1387) for general audio \cite{itubs1387} predict overall perceived quality relative to a reference signal; they do not assess whether the specific musical features required by a receiving task are preserved, and task-specific subjective assessment requires separate evaluation. We propose candidate diagnostics (Table~\ref{tab:metrics}), not validated quality-of-experience instruments. Fix features, matching, weights, tolerances, and observation points before comparing systems.

\textbf{Semantic Musical Fidelity (SMF)} compares a task reference $S_k$ and receiver representation $R_k$ for reference events/windows $k$, using $J$ required features:
\begin{equation}
\mathrm{SMF}_k=1-\sum_{j=1}^{J}w_j\delta_j(S_k,R_k)\in[0,1].
\label{eq:smf}
\end{equation}
Use $w_j\geq0$, $\sum_jw_j=1$, and declared losses $\delta_j\in[0,1]$, e.g., $\min\{|S_{k,j}-R_{k,j}|/\eta_j,1\}$ with tolerance $\eta_j>0$ in matching units, or categorical mismatch. Missing/invalid required receiver features incur loss one without weight renormalization. Fix a nonempty valid reference set and feature applicability independently of receiver outcomes.
Report the mean of $\mathrm{SMF}_k$ over the reference set, together with component errors, missingness, and extra events.
Decoded-descriptor comparisons measure descriptor fidelity; rendered-feature fidelity requires re-extraction from outputs. Neither validates reference extraction or perceived quality.

\textbf{Ensemble Synchronization Index (ESI)} combines output coverage and relative alignment. Endpoint $a$ denotes a receiver site and modality. Fix expected comparisons $\mathcal{P}=\{(k,a,b)\}$, counting each required unordered pair once per reference event, including task-relevant cross-site and cross-modal pairs. Declare event matching and output landmarks, not arrival-order matching. For observed output times $y_{a,k}$ and independently specified targets $p_{a,k}$ in common physical time,
\begin{equation}
r_{abk}=\left|(y_{a,k}-y_{b,k})-(p_{a,k}-p_{b,k})\right|.
\label{eq:esi-residual}
\end{equation}
Targets preserve expressive/modality offsets; they cannot come solely from the evaluated estimator. Musical tags are not observed times. With $Q=|\mathcal{P}|>0$,
\begin{equation}
\begin{aligned}
\mathrm{ESI}&=\frac{1}{Q}\sum_{(k,a,b)\in\mathcal{P}}q_{abk},\\
q_{abk}&=\begin{cases}
1-\min\{r_{abk}/\tau_{abk},1\},&\text{both times valid},\\
0,&\text{otherwise},
\end{cases}
\end{aligned}
\label{eq:esi}
\end{equation}
Fix tolerances $\tau_{abk}>0$ in the same units, observation horizon, and timestamp uncertainty. Expected pairs have equal weight. ESI lies in $[0,1]$; $Q=0$ is not applicable. Missing/invalid times score zero without shrinking $Q$. Report pair coverage, valid residual mean/tails/maximum, misses, extra events, event-order errors, absolute target errors, and end-to-end latency. Common delay cancels; per-pair clipping and averaging hide extreme errors. ESI alone establishes neither timely delivery nor musical benefit.

\textbf{Emotional Coherence Score (ECS)} is an exploratory co-variation diagnostic. For one affect coordinate (e.g., valence), let $x_A,x_K,x_V$ denote audio, kinematic, and visual trajectories on a common grid/affect scale over segment $s$. With $\mathcal{C}=\{(A,K),(A,V),(K,V)\}$,
\begin{equation}
\mathrm{ECS}(s)=\frac{1}{3}\sum_{(m,n)\in\mathcal{C}}\rho_s(x_m,x_n),
\label{eq:ecs}
\end{equation}
where $\rho_s$ is Pearson correlation over time. Fix the coordinate, models, segment, and alignment; score coordinates separately. Require at least three common finite samples and nonzero variance in all trajectories. Missing/nonfinite samples or constant trajectories make ECS not applicable; report valid-segment coverage. Its range is $[-1/2,1]$ on common support. Positive affine changes preserve correlation despite differing affective levels. Task baselines must allow intended contrast; ECS establishes neither emotional agreement, source-intent preservation, nor human experience.

\begin{table}[t]
\caption{Candidate Task-Level Diagnostics and Their Scope}
\label{tab:metrics}
\centering
\begin{tabularx}{\columnwidth}{@{}p{0.12\columnwidth}p{0.38\columnwidth}Y@{}}
\toprule
\textbf{Metric} & \textbf{Question answered} & \textbf{Key design choice} \\
\midrule
SMF & Are declared features preserved? & Specify decoded or rendered features; fixed losses/weights. \\
ESI & Are expected output relationships preserved? & Independent timing targets; missing pairs score zero. \\
ECS & Do inferred affect trajectories co-vary? & Common-sample correlation, not affective agreement. \\
\bottomrule
\end{tabularx}
\end{table}
 
The diagnostics may overlap, e.g., when SMF includes timing. Their task-specific instantiation and perceptual validation remain research requirements.

\begin{figure*}[!t]
\centering
\includegraphics[width=\textwidth]{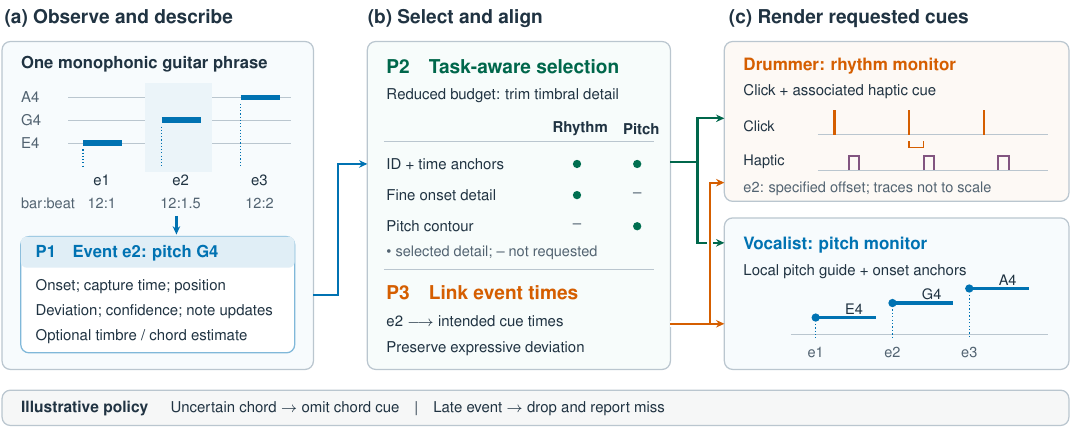}
\caption{Illustrative semantic flow for one guitar phrase, not a measured experiment or a wire-format specification. Positions use bar:beat notation; pitch segments and cue traces are schematic. Both recipients retain event identity and timing anchors. Dots mark selected fields; dashes mark unrequested detail. Under a reduced budget, optional timbral detail is reduced. P3 associates outputs with the same event and their intended relative offsets. The shown policy omits uncertain chord cues and declares late events missed; alternative policies need explicit evaluation.}
\label{fig:musical-flow}
\end{figure*}

\section{A Semantic-First Architecture}
Figure~\ref{fig:architecture} connects the three pillars through a shared musical event record. The placement principle is to execute each function at the tier closest to the participants that can support its latency and temporal-context requirements. Sender devices host P1, extracting local features and accumulating context; the edge hosts P2, combining receiver requests with participant and network state, together with the timing service of P3; receivers complete P3 by computing intended output times and rendering cues locally; and the cloud supports longer-horizon ($L_4$) inference and model adaptation outside the fast interaction loop. These are proposed roles, not claims of end-to-end implementation.

\subsection{Pillar 1 in Practice: Musical Semantic Extraction}

As an illustrative monitoring task within UC1 (remote jamming), consider the monophonic guitar phrase of Fig.~\ref{fig:musical-flow}(a). A device-side semantic encoder would expose onset, pitch, and optional timbral descriptors, updating the duration as the note unfolds, while accumulating past observations to estimate beat position or harmonic context. 
An event record associates these values with an identity, capture time, musical position $q_k$, expressive timing offset $\delta_k$, and confidence (Eq.~\ref{eq:scheduled-target}). A structural estimate is available only after sufficient evidence; its update rate must not be confused with its observation window or look-ahead.

The set of configurable semantic fields makes the representation task-dependent: a drummer can request finer onset detail, while a vocalist can request a pitch contour. Each requested field should satisfy the extractability, communicability and actionability criteria of Section~IV-A. Local renderers could turn those descriptors into monitoring cues; reconstruction of the original guitar timbre is a separate, unvalidated requirement. The architectural question is which named dimensions, resolutions, and confidence information each musical task requires.

\subsection{Pillar 2 in Practice: Musical Priority Routing}

Figure~\ref{fig:musical-flow} follows event e2 through the proposed interface. Its onset and pitch G4 are associated with the same event identity. For each receiver, the edge router selects fields and resolutions from the event record according to the receiver's request, the confidence of each field, and the available budget. When bandwidth falls, it could reduce timbral detail while retaining finer onset information for the drummer's rhythm monitor and pitch-contour detail for the vocalist's pitch monitor; both branches retain the event identity and timing anchors. The receiver actions are a click with an associated haptic cue and a local pitch guide, respectively; these are illustrative functions, not demonstrated improvements in ensemble performance.

Priorities follow the requested task rather than the role name alone: a sound engineer assessing timbre may require exactly the detail these monitor functions omit. Confidence also constrains selection: an uncertain chord estimate should not force a harmonic decision, so the example omits that cue while retaining valid local features. Conventional priority queues can execute the chosen policy; the proposed interface supplies the musical dimensions, requests, and confidence on which that policy acts.

\begin{figure*}
\centering
\includegraphics[width=0.8\textwidth]{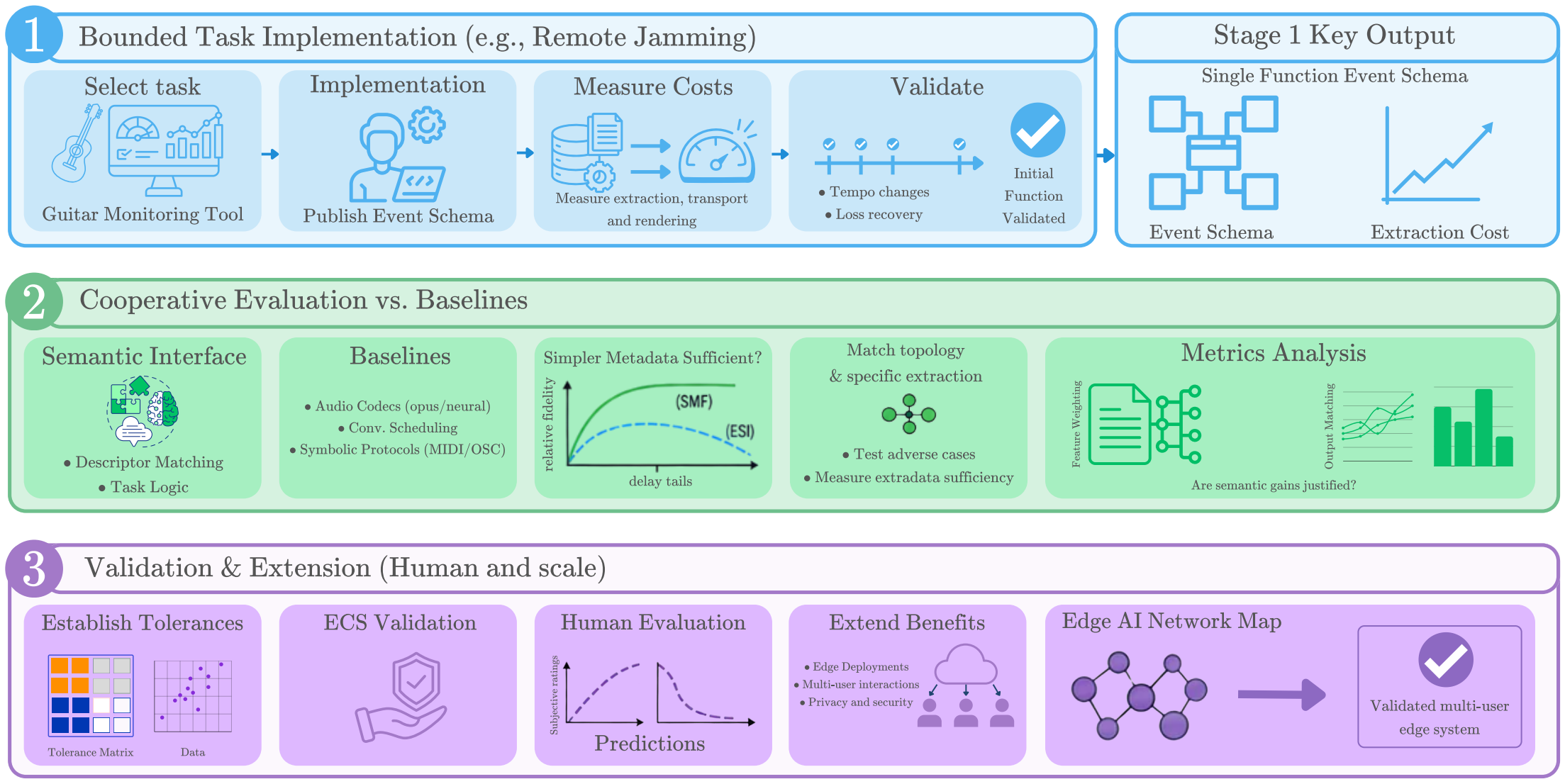}
\caption{General Design of a research roadmap for musical semantic communication, progressing from bounded task implementation to baseline comparison, metric-driven evaluation, and human and multi-user edge validation.}
\end{figure*}

\subsection{Pillar 3 in Practice: Semantic Synchronization}

Synchronization integrates shared musical identity and intended output relationships with conventional clock discipline and scheduled playback. In addition to the fields of Section~V-A, the event record carries a session/epoch, the source availability time, provenance (observed, announced, or predicted), and the required output landmarks. Observed content is available only after capture/encoding; prediction requires separate error accounting.

A designated timing service supplies a positive-tempo, piecewise-affine beat-to-reference-clock map $\widehat G_v$, with version $v$, generation time, effective beat/time, and expiry. Receivers accept only received, valid updates in increasing version order within the epoch. Effective intervals exclude retroactive changes; future-effective versions do not erase still-applicable segments. For endpoint $a$, the requested local target is
\begin{equation}
\widehat\ell_{a,k}=\widehat C_a\!\left(\widehat G_v(q_k)+\delta_k+D+o_a\right),
\label{eq:scheduled-target}
\end{equation}
where $\widehat C_a$ converts reference time to local clock time, and the expressive offset $\delta_k$, the declared playout allowance $D$, and the intended modality-landmark offset $o_a$ are in reference-clock seconds. One applicable received map snapshot is used per event. Admission requires
\begin{equation}
u_{a,k}+c_{a,k}+\epsilon_a\leq\widehat\ell_{a,k},
\label{eq:readiness}
\end{equation}
where $u$ is local content-availability time, $c$ bounds remaining preparation, and $\epsilon_a$ reserves clock/output uncertainty, in local time units. Timeline confidence, age, and error limits are declared separately. These conditional design requirements establish no measured precision guarantee. For event e2, the receiver maps its beat position to reference time, adds its declared expressive deviation, adds the playout and modality offsets, and converts the result to its local clock. The audio and haptic endpoints admit their cues only if each can prepare its output before that requested target.

The illustrative policy freezes accepted targets, suppresses duplicate epoch/event/endpoint realizations, and rejects targets violating declared within-voice order against queued or completed output.   If no valid map is available, the receiver falls back to source-landmark timestamp scheduling, but only with a valid clock relation and under the same allowance, offsets, and admission rules; otherwise, the event is dropped. Readiness failures are misses, not events shifted to another beat. State updates cannot replay committed output.

Evaluation uses observed output times and independent ESI targets $p$, not $\widehat\ell$, alongside absolute errors, latency, coverage, and order diagnostics. Estimator design, state recovery, and renderer behavior require implementation and comparison with equally informed timestamp scheduling under matched budgets. Neither the contract nor the figures establish real-time edge feasibility.

\section{Research Roadmap and Conclusion}

The framework connects musical requirements, descriptor choices, and output relationships through a common event interface. Its contribution is a reference architecture, taxonomy, worked example, and candidate diagnostics. The following stages turn that vision into falsifiable research questions.

\textbf{First, specify and implement a bounded task.} Select one monitoring or production function and publish its event schema, feature losses, timing landmarks, confidence policy, and failure rules. Measure capture/extraction, transport, control traffic, and rendering costs. Test causal availability and state recovery under tempo changes, clock error, loss, and receiver restarts before claiming deadline compliance.

\textbf{Second, test the value of the musical interface.} Compare against suitable audio codecs, symbolic/event protocols, and conventional priority scheduling. Match topology, multicast support, available information, look-ahead, buffering, and compute/wire budgets. Include descriptor errors and omitted musical detail, not only successful cue alignment. Report independent task outcomes, rate/fidelity trade-offs, coverage, delay tails, and adverse cases alongside SMF/ESI; test whether simpler metadata-aware baselines achieve the same result.

\textbf{Third, validate with musicians and heterogeneous implementations.} Establish task-specific tolerances and whether the diagnostics predict performer or audience judgments. ECS needs separate validation of affect inference and intended contrast. Extend only supported benefits to multi-user edge deployments, interoperability, and privacy/security evaluation.


\end{document}